\documentclass[12pt]{JHEP} % 10pt is ignored!
\usepackage{graphicx}
\newcommand\fverb{\setbox\pippobox=\hbox\bgroup\verb}
\newcommand\fverbdo{\egroup\medskip\noindent%
			\fbox{\unhbox\pippobox}\ }
\newcommand\fverbit{\egroup\item[\fbox{\unhbox\pippobox}]}
\newbox\pippobox

\def\ol{\overline}

\def\del{\partial}

\def\gap#1{\vspace{#1 ex}}

\def\eps{{\epsilon}}

\def\be{\begin{equation}}
\def\ee{\end{equation}}
\def\ba{\begin{array}{l}}
\def\ea{\end{array}}
\def\bea{\begin{eqnarray}}
\def\eea{\end{eqnarray}}
\def\beas{\begin{eqnarray*}}
\def\eeas{\end{eqnarray*}}
\def\eq#1{(\ref{#1})}

\def\nn{\nonumber\\}

\def\gap#1{\vspace{#1 ex}}

\def\ra{\rightarrow}
\def\eps{{\epsilon}}
\def\db{$D8$-brane}
\def\dbar{$\ol{D8}$-brane}

\title{Sakai-Sugimoto model, Tachyon Condensation and 
Chiral symmetry Breaking}
\author{Avinash Dhar and Partha Nag \\  
Tata Institute of Fundamental Research, Homi Bhabha Road, \\
Mumbai 400 005, India 
\\~~\\
\email{adhar@theory.tifr.res.in, parthanag@theory.tifr.res.in
}}
\preprint{TIFR/TH/07-19}
\abstract
{We modify the Sakai-Sugimoto model of chiral symmetry breaking to
take into account the open string tachyon which stretches between the
flavour \db s and \dbar s. There are several reasons of consistency
for doing this: (i) Even if it might be reasonable to ignore the
tachyon in the ultraviolet where the flavour branes and antibranes are
well separated and the tachyon is small, it is likely to condense and
acquire large values in the infrared where the branes meet. This takes
the system far away from the perturbatively stable minimum of the
Sakai-Sugimoto model; (ii) The bifundamental coupling of the tachyon
to fermions of opposite chirality makes it a suitable candidate for
the quark mass and chiral condensate parameters. We show that the
modified Sakai-Sugimoto model with the tachyon present has a classical
solution satisfying all the desired consistency properties. In this
solution chiral symmetry breaking coincides with tachyon
condensation. We identify the parameters corresponding to the quark
mass and the chiral condensate and also briefly discuss the mesonic
spectra.}


\begin{document}  
    
%%%%%%%%%%%%%%%%%%%%%%%%%%%%%%%%%%%%%%%%%%%%%%%%%%%%%%%%%%%%%%%%%%%%%%%%

\section{Introduction}

The study of connections between gauge theory and string theory in the
last decade, following the AdS/CFT conjecture \cite{JM,GKP}, has led
to the development of new tools for investigating strong coupling
phenomena in gauge theories \cite{EW1,EW2,JPS,KS,MN}. These
`holographic methods' have been used with surprising success in
qualitative studies of confinement and chiral symmetry breaking in
realistic QCD-like gauge theories, although application to real QCD,
which requires quantizing strings moving on highly curved spaces in
the presence of RR backgrounds, is still beyond the currently
available tools.

In the context of these holographic methods, a subject that has
received a lot of attention recently is that of chiral symmetry
breaking in QCD-like gauge theories. In holographic models of gauge
theories, the Yang-Mills fields arise from massless open string
fluctuations of a stack of `colour' branes. The near horizon, strong
coupling limit of a large number $N_c$ of colour branes has a dual
description in terms of a classical gravity theory. Flavour degrees of
freedom are introduced in this setting as the fermionic open string
fluctuations between the colour branes and an additional set of
`flavour' branes \cite{KK,BEEGK,KMMW,SJS1,BHMM,HN}. In the probe
approximation in which the number of flavour branes, $N_f$, remains
finite as $N_c \ra \infty$, the backreaction of the flavour branes on
the background geometry can be neglected and various phenomena
associated with flavour physics studied as classical effects in the
background geometry.

The model of Sakai and Sugimoto \cite{SS1}, which is based on this
scenario, has been very successful in reproducing many of the
qualitative features of non-abelian chiral symmetry breaking in QCD.
In this model, chiral symmetry breaking has a nice geometrical
picture. In the ultraviolet, chiral symmetry arises on flavour \db s
and \dbar s, which are located at well-separated points on a circle,
while they are extended along the remaining eight spatial directions,
including the holographic radial direction. Chiral symmetry breaking
in the infrared is signalled by a smooth joining of the flavour branes
and antibranes at some point in the bulk. At finite temperatures,
chiral symmetry is restored at or above the deconfinement transition
\cite{ASY,AHJK,PS}.

Despite its many qualitative and, remarkably, some quantitative
successes \cite{SS1,SS2,HSSY,HRYY,NSK,BLL,DY}, the Sakai-Sugimoto model has
some deficiencies. As has been pointed out by many authors
\footnote{See, for example \cite{ASY,CPS,AHJK,PSZ}.}, this model does
not have parameters associated either with the chiral condensate or
with quark bare mass. In addition, the model ignores the open string
tachyon between \db~and \dbar, which may be reasonable in the
ultraviolet where the branes and antibranes are well separated, but is
not so in the infrared where the branes join \footnote{Actually, even
in the ultraviolet region this is not so straightforward, see the
discussion in \cite{AHJK}.}. In this region one would expect the
tachyon to condense. Since the tachyon field takes an infinitely large
value in the true ground state \footnote{For a recent review of this
subject, see \cite{AS1}.}, the perturbative stability argument given
in \cite{SS1}, valid for small fluctuations of the tachyon field near
the local minimum at the origin, does not apply. 

Recently, it was suggested in \cite{CKP} that tachyon condensation on
a coincident brane-antibrane configuration describes the physics of
chiral symmetry breaking in a better and more complete
way. Unfortunately in this scenario one loses the nice geometric
picture of the Sakai-Sugimoto model for non-abelian chiral symmetry
breaking. The aim of the present work is to develop a model which
retains the nice features of the Sakai-Sugimoto model while overcoming
its deficiencies. We argue that this can be done by taking into
account the open string tachyon that stretches between separated
\db s and \dbar s. We will show that in our model, chiral symmetry 
breaking, which is signalled by joining of branes and antibranes, is
accompanied by tachyon condensation, since the tachyon field takes
large values only in the region where the branes and antibranes
join. Furthermore, the tachyon profile provides the necessary
parameters to describe both the quark mass and the chiral condensate.

The organization of this paper is as follows. In the next section we
will briefly review the essential features of the Sakai-Sugimoto
model. In section \ref{tachyon} we desribe the modification in this
model required to include the open string tachyon between the \db s
and \dbar s. We compute the contribution to the bulk energy momentum
tensor of this system and verify that the backreaction is small
everywhere. In this section we also obtain the classical solution for
the brane profile and the tachyon and identify the parameters
associated with the quark mass and the chiral condensate. Mesonic
fluctuations around this classical solution are briefly discussed in
section \ref{meson}. We end with a discussion in section
\ref{discussion}.

As this work was nearing completion, the paper \cite{BSS} appeared on
the archive which also discusses similar issues.

\section{\label{ss}The Sakai-Sugimoto model}

The Yang-Mills part of this model is provided by the near horizon
limit of a set of $N_c$ overlapping $D4$-branes, filling the
$(3+1)$-dimensional space-time directions $x^\mu$
$(\mu=1,2,3~\rm{and}~0)$ and wrapping a circle in the $x^4$ direction
of radius $R_k$, with antiperiodic boundary condition for fermions,
which gives masses to all fermions at the tree level (and scalars at
one-loop level) and breaks all supersymmetries. At low energies, the
theory on the $D4$-branes is $(4+1)$-dimensional pure Yang-Mills with
't Hooft coupling $\lambda_5=(2\pi)^2 g_s l_s N_c$ of length dimension. At
energies lower than the Kaluza-Klein scale $1/R_k$, this reduces to
pure Yang-Mills in $(3+1)$ dimensions. This is true in the weak
coupling regime, $\lambda_5 << R_k$, in which the dimensionally
transmuted scale developed in the effective Yang-Mills theory in
$(3+1)$ dimensions is much smaller than the Kaluza-Klein scale, which
is the high energy cut-off for the effective theory. In the strong
coupling regime, $\lambda_5 >> R_k$, in which the dual gravity
description is reliable, these two scales are similar. Therefore in
this regime there is no separation between the masses of glueballs and
Kaluza-Klein states. This is one of the reasons why the gravity regime
does not describe real QCD, but the belief is that qualitative
features of QCD like confinement and chiral symmetry breaking, which
are easy to study in the strong coupling regime, survive tuning of the
dimensionless parameter $\lambda_5/R_k$ to low values.

Sakai and Sugimoto introduced flavours in this setting by placing a
stack of $N_f$ overlapping \db s at the point $x^4_L$ and $N_f$
\dbar s at the point $x^4_R$ on the thermal circle. Massless open
strings between $D4$-branes and \db s, which are confined to the
$(3+1)$-dimensional space-time intersection of the branes, provide
$N_f$ left-handed flavours. Similarly, massless open strings between
$D4$-branes and \dbar s provide an equal number of right-handed
flavours, leading to a global $U(N_f)_L \times U(N_f)_R$ chiral
symmetry. This global chiral symmetry is visible on the $D8$ and
\dbar s as chiral gauge symmetry. 

In the large $N_c$ and strong coupling limit the appropriate
description of the wrapped $D4$-branes is given by the dual background
geometry. This background solution can be obtained from the type IIA
sugra solution for non-extremal $D4$-branes by a wick rotation of one
of the four noncompact directions which the $D4$-branes
fill, in addition to wrapping the compact (temperature) direction. In
the near horizon limit, it is given by \cite{EW2,IMSY}
\bea
ds^2&=&\left(\frac{U}{R}\right)^{3/2}\left(\eta_{\mu \nu}dx^{\mu}dx^{\nu}+
f(U)~(dx^4)^2 \right)+ \left(\frac{R}{U}\right)^{3/2}\left(\frac{dU^2}
{f(U)}+U^2d\Omega_4^2\right), \nn
e^{\phi}&=&g_s\left(\frac{U}{R}\right)^{3/4},
\qquad \qquad F_4=\frac{2\pi N_c}{V_4}\eps_4, \qquad \qquad 
f(U)=1-\frac{U_{k}^3}{U^3}, 
\label{bgd}
\eea
where $\eta_{\mu \nu}=\rm{diag}(-1,+1,+1,+1)$ and $U_k$ is a constant
parameter of the solution. $R$ is related to the $5$-d Yang-Mills
coupling by $R^3=\frac{\lambda_5 \alpha'}{4\pi}$. Also,
$d\Omega_4,~\eps_4$ and $V_4=8\pi^2/3$ are respectively the line
element, the volume form and the volume of a unit $S^4$.

The above metric has a conical sigularity at $U=U_k$ in the $U-x^4$
subspace which can be avoided only if $x^4$ has a specific
periodicity. This condition relates the radius of the circle in the
$x^4$ direction to the parameters of the background by
\be
R_k=\frac{2}{3} \left(\frac{R^3}{U_k}\right)^{\frac{1}{2}}
\label{kk}
\ee
For $\lambda_5 >> R_k$ the curvature is small everywhere and so the
approximation to a classical gravity background is reliable. As
discussed in \cite{IMSY}, at very large values of $U$, the string coupling
becomes large and one has to lift the background over to the
$11$-dimensional M-theory description.

Now consider a set of $N_f$ $D8$-\dbar~pairs in the above background,
placed at points $x^4_L=l/2$ and $x^4_R=-l/2$ respectively on the
circle. If $N_f$ is kept fixed as the large $N_c$ limit is taken, the
effect of the flavour branes on the background geometry should be
small and may be treated in the probe approximation. For the simple
case of a single $D8$-\dbar~pair, the action is
$$S=-\mu_8 \int d^9\sigma~e^{-\phi}
\left(\sqrt{-\rm{det}~A_L}+\sqrt{-\rm{det}~A_R}~\right),$$
where $\mu_8=1/(2\pi)^8l_s^9$ and $(A_{L,R})_{ab}=g_{MN}\del_a x_{L,R}^M
\del_b x_{L,R}^N$ is the induced metric on the brane. The indices
$a,b$ run over the world-volume directions of the branes while the
indices $M,N$ run over the background ten-dimensional space-time
directions. Using the static gauge and assuming $l$ depends on $U$
only, the action becomes 
$$S=-{\cal T}_8 V_4 \int d^4x \int dU
\left(\frac{U}{R}\right)^{-3/4} U^4 \left( \sqrt{D_L} +
\sqrt{D_R}~\right),$$
where ${\cal T}_8=\mu_8/g_s$ is the $D8$-brane tension and 
\be
D_L=D_R \equiv D=f(U)^{-1}\left(\frac{U}{R}\right)^{-3/2}+f(U)
\left(\frac{U}{R}\right)^{3/2}\frac{{l'(U)}^2}{4}.
\label{d}
\ee
Here and in the following a prime denotes derivative with respect to $U$. 

In the above setting chiral symmetry breaking has a geometrical
description. It is signaled by the brane-antibrane meeting at an
interior point $U \geq U_k$, even when they are well separated
asymptotically. This is because in the background geometry \eq{bgd}
the branes have nowhere to end and hence they must meet. This can also
be seen by explicitly solving the equation of motion for $l(U)$
obtained from the above action. This equation is 
\be
\left(\frac{\left(\frac{U}{R}\right)^{13/4}}{\sqrt{D}}\frac{f(U)}{4}
\left(\frac{U}{R}\right)^{3/2}l'(U)\right)'=0,
\label{no-t0}
\ee
which has the solution
\be
\frac{l(U)}{2}=U_0^4 f(U_0)^{1/2} \int_{U_0}^U dy 
\frac{f(y)^{-1}\left(\frac{y}{R}\right)^{-3/2}}{\sqrt{y^8f(y)-U_0^8f(U_0)}}.
\label{no-t1}
\ee
The branes meet at the point $U=U_0$, so $l(U_0)=0$. Moreover, the
solution determines the asymptotic separation $l_0$ of the branes in
terms of $U_0$. The case in which there is maximum separation between
the brane and antibrane, $l_0=\pi R_k$, is special since in this case
$l(U)$ is independent of $U$.

In the generic case, the brane-antibrane system looks like a single
brane, coming in from the asymptotic region, turning around near
$U=U_0$ and returning back to the position of the other brane in the
asymptotic region. Expanding around the point $U=U_0$, we get from
\eq{no-t1}
\be
\frac{l(U)}{2}=\frac{R^{3/2}}{U_0 \sqrt{f(U_0)}} \frac{(U-U_0)^{1/2}}
{\sqrt{3+5f(U_0)}}[1+O(U-U_0)]. 
\label{no-t2}
\ee
We see that $l'(U) \sim (U-U_0)^{-1/2}$ diverges near the turning
point of the brane profile, as required by a smooth joining of the
brane with the anti-brane.

\section{\label{tachyon}Sakai-Sugimoto with tachyon}

The effective field theory describing the dynamics of a
brane-antibrane pair \footnote{For simplicity, we will continue to
discuss the case of a single flavour, namely one brane-antibrane
pair. Generalization to the multi-flavour case can be done using the
symmmetrized trace prescription of \cite{AT}.} with the tachyon
included has been discussed in \cite{AS2,MG}. The simplest case occurs
when the brane and antibrane are on top of each other since in this
case all the transverse scalars are set to zero. This is the situation
considered in \cite{CKP}. However, in this configuration one loses the
nice geometrical picture of chiral symmetry breaking of the
Sakai-Sugimoto model. Since we would like to retain this geometrical
picture, we must consider the case when the brane and antibrane are
separated in the compact $x^4$ direction. This requires construction
of an effective tachyon action on a brane-antibrane pair, taking into
account the transverse scalars. Such an effective action with the
brane and antibrane separated along a noncompact direction has been
proposed in \cite{AS2,MG}. A generalization of this action to the
present case when the brane and antibrane are separated along a
periodic direction is not known. However, for small separation
compared to the radius of the circle, the action in
\cite{MG} should provide a reasonable approximation. In the following
we will assume this to be the case. Then, the effective tachyon action
for $l(U) << R_k$ is
\bea
S &=& -\int d^9\sigma~V(T,l)e^{-\phi}
\left(\sqrt{-\rm{det}~A_L}+\sqrt{-\rm{det}~A_R}~\right), \nn
(A_{i})_{ab} &=& \left(g_{MN}-\frac{T^2l^2}{Q}g_{M4}g_{4N}\right)
\del_ax^M_i \del_bx^N_i+F^i_{ab} + \frac{1}{2Q}\biggl((D_a\tau (D_b\tau)^*
+(D_a\tau)^* D_b\tau) \nn
&& +il(g_{a4}+\del_ax^4_i g_{44})
(\tau (D_b\tau)^*-\tau^* D_b\tau)+il(\tau (D_a\tau)^*-\tau^* D_a\tau)
(g_{4b}-\del_bx^4_i g_{44})\biggr), \nn
\label{with-t1}
\eea
where
\be
Q=1+T^2l^2g_{44}, \quad 
D_a\tau=\del_a \tau-i(A_{L,a}-A_{R,a})\tau, \quad V(T,l)=g_s V(T)\sqrt{Q}.
\label{with-t2}
\ee
$T=|\tau|$, $i=L,R$ and we have used the fact that the background does
not depend on $x^4$. Also, in writing the above we are using the
convention $2\pi \alpha'=1$. \footnote{The complete action also
includes Chern-Simons (CS) couplings of the gauge fields and the
tachyon to the RR background sourced by the $D4$-branes.These will not
be needed in the following analysis and hence have not been included
here.}

The potential $V(T)$ depends only on the modulus $T$ of the complex
tachyon $\tau$. It is believed that $V(T)$ satisfies the following
general properties \cite{AS1}:
\begin{itemize}
\item
$V(T)$ has a maximum at $T=0$ with $V(0)={\cal T}_8$.
\item
The normalization of $V(T)$ is fixed by the requirement that the
vortex solution on the brane-antibrane system produce the correct
relation between $Dp$ and $D(p-2)$-brane tensions.
\item
In flat space for brane-antibrane on top of each other (i.e. for
$l=0$), the expansion of $V(T)$ around $T=0$ upto terms quadratic in
$T$ gives rise to a tachyon with mass-squared equal to $-\pi$ in our
conventions.
\item
$V(T)$ has a minimum at $T=\infty$ where it vanishes.
\end{itemize}

There are several proposals for $V(T)$ which satisfy these
requirements \cite{AS1}, although no rigorous derivation exists. In
view of this, in the following analysis we will avoid using any
specific expression for $V(T)$, except when needed for explicit
numerical calculations. It will, however, be necessary for us to
specify the asymptotic form of the potential for large $T$. We will
assume that in our parametrization this behaviour is given by $V(T)
\sim e^{-cT}$ where $c$ is a positive constant. A potential satisfying
this property, in addition to the properties listed above is
\cite{KKKK,LP,LLM}
\be
V(T)=\frac{{\cal T}_8}{{\rm cosh}{\sqrt \pi}T}.
\label{pot}
\ee

\subsection{\label{emt}Backreaction of the flavour branes}

Let us now first discuss the backreaction on the background
geometry. For this we need to compute the contribution of the flavour
brane-antibrane system to the ten-dimensional bulk energy momentum
tensor. Our starting point is the action \eq{with-t1}. The energy
momentum tensor is obtained from it by calculating its functional
derivative w.r.t. the background ten-dimensional metric $g_{MN}$. The
precise relation is $T^{MN}=2/\sqrt{-{\rm det}~g}~\delta S/\delta
g_{MN}.$ We get,
\bea
T_i^{ab} &=& -g_s V(T) \sqrt{Q} e^{-\phi} \frac{\sqrt{-{\rm det}~A_i}}
{\sqrt{-{\rm det}~g}} \left(A_i^{-1}\right)^{ab}_{S}, \nn
T_i^{a4} &=& -g_s V(T) \sqrt{Q} e^{-\phi} \frac{\sqrt{-{\rm det}~A_i}}
{\sqrt{-{\rm det}~g}}~2\left(A_i^{-1}\right)^{ab}_{S}
\left(\del_bx_i^4-T^2 l A_b\right), \nn
T_i^{44} &=& -g_s V(T) \frac{1}{\sqrt{Q}} e^{-\phi} 
\frac{\sqrt{-{\rm det}~A_i}}{\sqrt{-{\rm det}~g}} \times \nn
&& \left[-8T^2l^2+\left(A_i^{-1}\right)^{ba}
\left(T^2l^2(g_{ab}+F_{ab})+\del_ax_i^4\del_bx_i^4
+T^2l(A_a \del_bx_i^4- a \leftrightarrow b)\right)\right], \nn
\label{emt1}
\eea
where $i=L$ ($R$) denotes the contribution of the \db~(\dbar) and the
subscript `$S$' stands for the symmetric part. Also, we have defined
$A_b \equiv (A_{Lb}-A_{Rb}-\del_a\theta)$, where $\theta$ is
the phase of the complex tachyon, $\tau=Te^{i\theta}$. It is
understood that each of the above expressions must be multiplied by a
position space delta-function specifying the location of the brane in
the transverse space where its contribution to the ten-dimensional
bulk energy momentum tensor is localized.

Specializing these expressions to the case of the background solution
where the gauge fields are set to zero and $T$ and $l$ are functions
of $U$ only, we get
\bea
T_i^{ab} &=& -V(T) \left(\frac{U}{R}\right)^{-3/4} 
\sqrt{D_T}~g^{ab}, \quad a,b \neq U, \nn
T_i^{UU} &=& -V(T) \left(\frac{U}{R}\right)^{-3/4} 
\frac{Q}{\sqrt{D_T}}, \nn
T_i^{44} &=& -V(T) \left(\frac{U}{R}\right)^{-3/4}
\frac{f^{-1}\left(\frac{U}{R}\right)^{-3/2}}{\sqrt{D_T}}
\left(T^2l^2+f\left(\frac{U}{R}\right)^{3/2}\frac{{l'}^2}{4}\right),
\label{emt2}
\eea
and all other components vanish. The quantity $D_T$ is defined in
\eq{bgd-t2}. If $T$ goes to infinity near the place where the brane 
and the antibrane meet, all the components of the energy momentum
tensor vanish there because $V(T) \rightarrow 0$ exponentially for
large values of $T$. Thus the situation is even better than without
the tachyon \footnote{In the absence of the tachyon, the energy
momentum tensor components in \eq{emt2} blow up near the place where
the brane and the antibrane meet. This is, however, not a real
singularity since it can be removed by changing the descrption, for
example, to $U$ as a function of $l$ instead of the description in
terms of $l(U)$.} and the flavour contribution to the energy momentum
tensor is small everywhere, justifying the probe approximation for a
generic configuration.

Recently a detailed calculation of the backreation of the flavour
branes on the geometry in the Sakai-Sugimoto model has been reported
in \cite{BKS}. In this work the calculation has been done for the
special configuration in which the branes and antibranes are separated
maximally on the circle, i.e. $l=\pi R_k$. The authors find that, as
expected, in this antipodal case the corrections are indeed small for
$N_f/N_c$ small. It would be interesting to extend their calculation
to the generic case with the tachyon present.

\subsection{\label{bgd-t}Tachyon condensation as chiral symmetry breaking}

We will now look for an appropriate classical solution of the
brane-antibrane-tachyon system. Let us set the gauge fields and all
but the derivatives with respect to $U$ of $T$ and $x_i^4$ to
zero. Moreover, we choose $x_L^4=l/2$ and $x_R^4=-l/2$ so that the
separation between the brane and antibrane is $l$. In this case, in
the static gauge the action
\eq{with-t1} simplifies to \footnote{The CS term in the action does
not contribute for such configurations.}
\be
S=-V_4 \int d^4x \int dU~V(T) \left(\frac{U}{R}\right)^{-3/4} U^4
\left(\sqrt{D_{L,T}}+\sqrt{D_{R,T}}~\right), 
\label{bgd-T1}
\ee
where $D_{L,T}=D_{R,T} \equiv D_T$ and
\be
D_T=f(U)^{-1}\left(\frac{U}{R}\right)^{-3/2}+f(U)
\left(\frac{U}{R}\right)^{3/2}\frac{{l'(U)}^2}{4}+{T'(U)}^2+T(U)^2l(U)^2.
\label{bgd-t2}
\ee
The equations of motion obtained from this action are
\bea
\left(\frac{U^{\frac{13}{4}}}{\sqrt{D_T}} T'(U)\right)' 
&=& \frac{U^{\frac{13}{4}}}{\sqrt{D_T}} 
\left[T(U)l(U)^2+\frac{V'(T)}{V(T)}(D_T-T'(U)^2)\right],
\label{eq-t} \\
\left(\frac{U^{\frac{13}{4}}}{\sqrt{D_T}}\frac{f(U)}{4}
\left(\frac{U}{R}\right)^{\frac{3}{2}}l'(U)\right)' 
&=& \frac{U^{\frac{13}{4}}}{\sqrt{D_T}}
\left[T(U)^2 l(U)-\frac{V'(T)}{V(T)}\frac{f(U)}{4}
\left(\frac{U}{R}\right)^{\frac{3}{2}}l'(U) T'(U)\right]. \nn
\label{eq-l}
\eea
Note that the `prime' on $V(T)$ denotes a derivative w.r.t. its
argument $T$ and not a derivative w.r.t. $U$.

This is a complicated set of coupled nonlinear differential
equations. To get some insight into the kind of solutions that are
possible, we will first analyse the equations for large $U$ and for
$U$ near the brane-antibrane joining point, where the equations
simplify and can be treated analytically. As in the case without the
tachyon, we are looking for solutions in which the brane and antibrane
have an asymptotic separation $l_0$, i.e. $l(U) \ra l_0$ as $U \ra
\infty$ and they join at some interior point in the bulk, i.e. $l(U)
\ra 0$ at $U=U_0 > U_k$. Moreover, we want the tachyon (i) to vanish
as $U \ra \infty$ so that the chiral symmetry is intact in the
ultraviolet region and (ii) to go to infinity as $U$ approaches $U_0$
so as to reproduce correctly the QCD chiral anomalies \cite{CKP}.

\subsubsection{\label{large-U}Solution for large $U$}

We are looking for a solution in which $l(U)$ approaches a constant
$l_0$ and $T$ becomes small as $U \ra \infty$. Let us first consider
the equation \eq{eq-t}. For small $T$ one can approximate $V'/V \sim
-\pi T$ \footnote{This follows from the general properties of the
potential discussed in section \ref{tachyon}.}. If $T$ and $l'$ go to
zero sufficiently fast as $U \ra \infty$ such that to the leading
order one might approximate $D_T \sim
\left(\frac{U}{R}\right)^{-3/2}$, then \eq{eq-t} reduces to
\be
\left(U^4~T'(U)\right)'=l_0^2~U^4~T.
\label{asym}
\ee
This equation can be solved exactly with the general solution
\be
T(U)=\frac{T_+}{U^2} (1+\frac{1}{l_0 U}) e^{-l_0 U}+
\frac{T_-}{U^2} (1-\frac{1}{l_0 U}) e^{l_0 U},
\label{sol-t}
\ee
The solution with the exponential fall off satisfies the
approximations under which \eq{asym} was derived for any large value
of $U$. The exponentially rising solution will, however, eventually
become large and cannot be self consistently used. This is because for
sufficiently large $U$, there is no consistent solution for $T$ which
grows exponentially or even as a power-law to the original equations
\eq{eq-t} and \eq{eq-l}, if we impose the restriction that
$l(U)$ should go to a constant asymptotically. This puts a restriction
on the value of $U$ beyond which the generic solution \eq{sol-t}
cannot be used. The most restrictive condition comes from the
approximation $D_T \sim
\left(\frac{U}{R}\right)^{-3/2}$. This requires the maximum value, 
$U_{\max}$, to satisfy $U_{\max}^{5/2}e^{-2l_0U_{\max}} >>
l_0^2T_-^2R^{-3/2}$.  At values of $U$ much larger than this, only the
exponentially falling part provides a consistent solution.

Even though \eq{sol-t} does not represent a truly asymptotic solution,
its usefulness lies in the fact that most quantities of interest that
involve the tachyon, like pseudoscalar meson masses, receive maximum
contribution from intermediate values of $U$ and hence from this
solution. This is because the exponentially falling tachyon potential
kills off contribution in the infrared region and the exponentially
falling tachyon does so in the ultraviolet region, so the maximum
contribution comes from intermediate region. Thus physical qantities
are sensitive to both the parameters of this solution. It is natural
to associate $T_-$ with the quark bare mass since this parameter comes
with the growing solution and $T_+$ with the chiral condensate because
it is associated with the normalizable solution. More evidence for
this will be given in the next section.

The fact that the tachyon takes small values for large $U$  makes it 
irrelevant for the
leading behaviour of $l$, which can be extracted from \eq{eq-l} by
setting the r.h.s. to zero. The resulting equation is precisely
\eq{no-t0} with a similar solution
\be
l(U)=l_0-l_1 U^{-9/2}+\cdots
\label{sol-l}
\ee
where $l_1$ is positive so that the branes come together. For 
Sakai-Sugimoto without the tachyon, $l_1=\frac{2}{3} R_k U_0^4 
\sqrt{U_k f_0}$, where $f_0=f(U_0)$.

Is there a solution in which $T$ vanishes asymptotically as a power
law? Suppose there is such a solution, $T(U) \sim U^{-\alpha}$. If
$\alpha > 3/4$ and $l$ vanishes fast enough, we may once again
approximate $D_T \sim \left(\frac{U}{R}\right)^{-3/2}$. As before, we
then conclude that $T$ vanishes exponentially, which contradicts our
assumption that $T$ vanishes as a power law. If $\alpha < 3/4$ and $l'$
vanishes fast enough, then we must approximate $D_T \sim
T^2l_0^2$. One can see immediately from \eq{eq-t} that this also leads
to a contradiction. Finally, suppose asymptotically $l'$ vanishes so
slowly that it is the $l'^2$ term that dominates in $D_T$ and so we
must approximate $D_T \sim (U/R)^{3/2}{l'(U)}^2/4$. Once again it is
easy to see from \eq{eq-t} that there is no consistent solution. We
thus conclude that the only solution in which $l$ goes to a nonzero
constant asymptotically and $T$ vanishes is the one given by \eq{sol-t},
\eq{sol-l} (after dropping the growing part of $T$ for large enough $U$).

\subsubsection{\label{small-U}Solution for $U \sim U_0$}

Here we are looking for a solution in which $l \ra 0$ and $T \ra
\infty$ as $U \ra U_0$. Let us assume a power law ansatz, namely
$$l(U) \sim (U-U_0)^\alpha, \quad \quad T(U) \sim (U-U_0)^{-\beta}.$$
For a smooth joining of the brane and antibrane at $U_0$, the
derivative of $l$ must diverge at this point, which is ensured if
$\alpha < 1$. Since for this ansatz $T'^2$ is the largest quantity for
$U \ra U_0$, we can approximate $D_T \sim T'(U)^2$. Moreover, using
the asymptotic form of the potential $V(T)
\sim e^{-cT}$ for large $T$, we get $V'(T)/V(T) \sim -c$. Putting all 
this in \eq{eq-t} we see that the leading term on the l.h.s. is a
constant. The first term on the r.h.s. vanishes as a positive power of
$(U-U_0)$. For consistency with the l.h.s. we then find from the
second term on the r.h.s. that (i) if $\beta > 1$, we must have
$\beta=1+2\alpha$ and (ii) if $\beta < 1$, we must have
$\beta=1-2\alpha$. $\beta=1$ is not allowed since we must have $0 <
\alpha < 1$. Analyzing equation \eq{eq-l} similarly, we find that in
case (i) the l.h.s. of this equation vanishes as a positive power of
$(U-U_0)$. This is consistent with the r.h.s. only if $\beta=2$, which
then gives $\alpha=1/2$. In case (ii) it is the first term on the
r.h.s. that vanishes as a positive power of $(U-U_0)$. Consistency
with the r.h.s. then requires $\beta=0$, which is however inconsistent
with our approximations. Hence, $\alpha=1/2,~\beta=2$ is the only
consistent solution we get which has $l \ra 0$ and $T \ra \infty$ as $U
\ra U_0$.  This ansatz can now be checked directly and the various
coefficients fixed. We get
\bea
l(U) &=& \frac{1}{v_1} \sqrt{\frac{26}{U_0f_0}} 
\left(\frac{U_0}{R}\right)^{-3/4}(U-U_0)^{1/2}+ \cdots,  \\
\label{sol-l5}
T(U) &=& \frac{v_1}{4} f_0 \left(\frac{U_0}{R}\right)^{3/2} (U-U_0)^{-2}
+ \cdots,
\label{sol-t1}
\eea
where $v_1$ is a constant which equals the limiting
value of $-V'(T)/V(T)$ as $T \ra \infty$. Note that, given the
potential, the normalizations of both $l$ and $T$ get fixed in terms
of $U_0$. It is important to mention that this solution exists only
for potentials which have the asymptotic behaviour $V(T) \sim
e^{-cT^\gamma}$ for large $T$, with $\gamma < 2$. \footnote{This
condition is not satisfied by the potential obtained by a boundary
string field theory computation
\cite{MZ,KMM,KL,TTU} for which $\gamma=2$. This is not necessarily a 
contradiction and probably indicates a nontrivial field redefinition
that relates fields we are using here to those used in the boundary
string field theory. A similar observation has been made earlier in
connection with the tachyon kink and vortex solutions on the
brane-antibrane system in \cite{AS2}. Note, however, that a calculation 
of S-matrix elements of tachyons and gauge fields reported in \cite{FG} 
seems to favour the boundary string field theory potential.}

The existence of the solution \eq{sol-t1}, \eq{sol-l5} shows that
tachyon condensation on the flavour brane-antibrane system is
intimately connected with chiral symmetry breaking.

For completeness, we note that there exists another solution in which
$T$ does not diverge as $U \ra U_0$. Let us assume that $T$ goes to a
nonzero constant as $U \ra U_0$. In this case we can approximate $D_T
\sim f(U)(U/R)^{3/2}{l'(U)}^2/4$. Substituting in \eq{eq-t} we see that 
the l.h.s. diverges as $(U-U_0)^{-\alpha}$. The first term on the
r.h.s. vanishes as a positive power, but the second term diverges as
$(U-U_0)^{\alpha-1}$. For consistency we must have $\alpha=1/2$. The
resulting solution 
\bea
l(U) &=& \frac{4}{U_0} \left(\frac{R^3}{f_0(5f_0+3)}\right)^{1/2}
(U-U_0)^{1/2}+ \cdots, \\
\label{sol-l2} 
T(U) &=& t_0+\frac{2}{(5f_0+3)} \left(\frac{R^3}{U_0}\right)^{1/2} 
\frac{V'(U_0)}{V(U_0)}(U-U_0)+ \cdots 
\label{sol-t2}
\eea
also satisfies \eq{eq-l}. Note that no special condition was required
for the tachyon potential to get this solution; this solution exists
for any potential.

To get a complete solution, one needs to use numerical tools since the
equations cannot be solved analytically. The numerical calculations
are in progress and will be reported in a forthcoming longer version
of this work \cite{DN}. 

%We have numerically integrated the equations using Mathematica and
%obtained a solution which has the above characteristics, namely
%unbroken chiral symmetry on separated branes in the ultraviolet region
%and broken chiral symmetry in the bulk signalled by the joining of
%brane and antibrane accompanied by a tachyon vev which goes to
%infinity at the place where they join. Figure 1 shows a plot for the
%numerical solution of the brane profile and Figure 2 shows the
%corresponding solution for the tachyon........MORE HERE.........
%\begin{figure}[htb] 
%\centering 
%\includegraphics[height=7cm,
%width=5cm]{1.eps} 
%\caption{Solution for the brane profile.}  
%\label{fig:1}
%\end{figure} 
%\begin{figure}[htb] 
%\centering 
%\includegraphics[height=7cm,
%width=5cm]{2.eps}
%\caption{Solution for the tachyon.}  
%\label{fig:2}
%\end{figure} 

\section{\label{meson}The meson spectra}

In this section we will discuss the spectra for various low spin
mesons which are described by the fluctuations of the flavour branes
around the classical solution. The action for the fluctuations of the
gauge fields can be computed from \eq{with-t1}. Parametrizing the
complex tachyon $\tau$ in terms of its magnitude and phase,
$\tau=Te^{i\theta}$, we get
\bea
\Delta S_{\rm gauge} &=& -\int d^4x~dU \biggl[a(U) A_U^2 + b(U) A_\mu^2 + 
c(U)\left((F^V_{\mu\nu})^2+(F^A_{\mu\nu})^2\right)+e(U) F^A_{\mu U} A^\mu \nn
&& + d(U) \left((F^V_{\mu U})^2+(F^A_{\mu U})^2\right) \biggr],
\label{gaugefluc1} \\
a(U) &=& V_4 V(T)U^4 \left(\frac{U}{R}\right)^{-3/4}
\frac{T^2}{\sqrt{D_T}},
\label{gaugefluc2} \\
b(U) &=& V_4 V(T)U^4 \left(\frac{U}{R}\right)^{-3/4} \sqrt{D_T} 
\left(\frac{U}{R}\right)^{-3/2} \frac{T^2}{Q}
\left(1+\frac{f^2T^2l^2l'^2}{4D_T}\left(\frac{U}{R}\right)^3\right), 
\label{gaugefluc3} \\
c(U) &=& V_4 V(T)U^4 \left(\frac{U}{R}\right)^{-3/4} \sqrt{D_T}
\frac{1}{8} \left(\frac{U}{R}\right)^{-3},
\label{gaugefluc4} \\
d(U) &=& V_4 V(T)U^4 \left(\frac{U}{R}\right)^{-3/4} 
\left(\frac{U}{R}\right)^{-3/2} \frac{Q}{4\sqrt{D_T}},
\label{gaugefluc5} \\
e(U) &=& V_4 V(T)U^4 \left(\frac{U}{R}\right)^{-3/4} 
\frac{fT^2ll'}{2\sqrt{D_T}}.
\label{gaugefluc5}
\eea
Here $F^V$ is the field strength for the vector gauge field
$V=(A_1+A_2)$ and $F^A$ is the field strength for the gauge-invariant
combination of the axial vector field and the phase of the tachyon,
$A=(A_1-A_2-\del\theta)$.

The gauge field $V_\mu(x,U)$ gives rise to a tower of vector mesons
while the fields $A_\mu(x,U)$ and $A(x,U)$, which are gauge invariant,
give rise to towers of axial and pseudoscalar mesons. Notice that the
coefficients $a(U)$, $b(U)$ and $e(U)$ vanish if the tachyon is set to
zero. In the absence of the tachyon the vector and axial vector mesons
acquire masses because of a nonzero $d(U)$, but there is always a
massless ``pion'' \footnote{Strictly speaking, for the $U(1)$ case
under discussion, this pseudoscalar is the $\eta'$. It is massless
here because of the $N_c \ra \infty$ limit in which we are
working.}. The presence of the tachyon is thus essential to give a
mass to the ``pion''. Also note that with the tachyon present, the
masses of the vector and axial vector mesons are in principle
different.

In the following, we will be using the gauge $V_U=0$. As we have
already noted, $A_U$ is gauge invariant. Expanding in modes, we have
\bea
V_\mu(x,U) &=& \sum_m V^{(m)}_\mu(x) W_m(U) \nn 
A_{\mu}(x,U)&=&\sum_m A^{(m)}_{\mu}(x) P_m(U),\nn
A_U(x,U)&=&\sum_m\phi^{(m)}(x)S_m(U),
\eea  
where $\{W_m(U)\}$, $\{P_m(U)\}$s and $\{S_m(U)\}$ form complete sets
of basis functions. The fields $\{V^{(m)}_\mu\}$, $\{A_{\mu}^{(m)}\}$ and
$\{\phi^{(m)}\}$ form towers of vector, axial-vector and pseudoscalar
mesons in the physical $(3+1)$-dimensional space-time. Note that
$\del^\mu A^{(m)}_\mu$ and $\phi^{(m)}$ mix. After suitably shifting
$A_{\mu}^{(m)}$ by an appropriate linear combination of
$\partial_{\mu}\phi^{(m)}$s, the mixing can be removed. The spectrum may
then be read off from the quadratic action
\bea
\Delta S_{\rm gauge}=&-&\int d^4x~\sum_m\biggl[\frac{1}{4} 
F^{V(m)}_{\mu\nu}F^{V(m)\mu\nu}+
\frac{1}{2} \lambda^V_m V^{(m)}_{\mu}V^{(m)\mu}+
\frac{1}{4} F^{A(m)}_{\mu\nu}F^{A(m)\mu\nu} \nn
&& +\frac{1}{2} \lambda^A_m A^{(m)}_{\mu}A^{(m)\mu}+
\frac{1}{2} \partial_{\mu}\phi^{(m)} 
\partial^{\mu}\phi^{(m)}
+\frac{1}{2} \lambda^\phi_m \phi^{(m)}\phi^{(m)}\biggr], 
\label{mesons}
\eea
where in the vector and axial vector sectors we have imposed the
orthonormality conditions
\bea
\int dU~c(U) P_m(U)P_n(U) &=& \frac{1}{4}\delta_{mn}
=\int dU~c(U) W_m(U)W_n(U), 
\eea
and the eigenvalue equations
\bea
-\left(d(U) W'_m(U)\right)' &=& 2 \lambda^V_m c(U) W_m, \nn
-\left(d(U)P'_m(U)\right)'+\left(b(U)+\frac{1}{2}e'(U)\right)P_m(U)
&=& 2 \lambda^A_m c(U) P_m(U).
\eea 
In the pseudoscalar sector, we need the conditions
\bea
\int dU~a(U) S_m(U)S_n(U) &=& \frac{1}{2} \lambda^\phi_m \delta_{mn}\nn
(K-\frac{1}{2} J^{\rm T}L^{-1}J)_{mn} &=& 
\frac{1}{2} \delta_{mn},
\label{pionmass}
\eea
where
\bea
J_{mn}&=& \int dU~[e(U)P_m(U)-2d(U)P_m^{\prime}(U)]S_n(U),\nn
K_{mn}&=& \int dU~d(U)S_m(U)S_n(U),\nn
L_{mn}&=& \lambda^A_m \delta_{mn}
\eea

One can also consider fluctuations in $T$ and $l$. There is mixing in
this sector also. These fluctuations give rise to towers of scalars
whose masses depend on the background value of the tachyon. We defer
details of these calculations to a forthcoming publication \cite{DN}.

\subsection{\label{GOR-relation}Quark mass and chiral condensate}

In this section we will give evidence for the identifications made
below \eq{sol-t} for the parameters $T_{\pm}$ with the chiral
condensate and quark mass. We note that for $T(U)=0$, $a(U)$ vanishes
and hence $\lambda^\phi_m$, given by the first of \eq{pionmass}, also
vanishes. We see once again that a nonzero tachyon is required for
nonzero pseudoscalar masses. Furthermore, since $V(T)$ vanishes
exponentially for large $T$, the region of $U$ in which $T$ is small,
but not too small, dominates the integral in \eq{pionmass}. This is
the intermediate region discussed below \eq{sol-t}. In this region $T$
can be essentially replaced by \eq{sol-t}. Consider the lightest mass
state. For this state, we have
\be
\frac{1}{2} \lambda^\phi_0=\int dU~a(U) (S_0(U))^2
\ee
The r.h.s. of this equation involves the quantity $a(U)$ which is
proportional to $T^2$. Using \eq{sol-t} and retaining to lowest order
in the quark mass parameter, which we have identified with $T_-$, we
see that this gives $\lambda^\phi_0 \sim T_-T_+$ \footnote{To get this
result, one first does the calculation with a given cut-off, $U_{\rm
max}$, with the condition $T_+ e^{-l_0 U_{\rm max}} >> T_- e^{l_0
U_{\rm max}}$. This is the condition that $T_-$ is small and justifies
retaining upto linear in $T_-$ term only. One must then remove the
cut-off, $U_{\rm max} \rightarrow \infty$, keeping the condensate and
the physical quark mass fixed. This should remove the $T_+^2$ term,
consistent with the fact that for $T_-=0$ the pion mass must
vanish.}. This firms up the identification of $T_+$ with the chiral
condensate. This relation is then essentially the
Gell-Mann-Oakes-Renner relation.

%\subsection{\label{vectors}Vector and Axial mesons}
%\subsection{\label{pions}Pseudoscalar mesons}
%\subsection{\label{scalars}Scalar mesons}

\section{\label{discussion}Discussion}

In this paper we have proposed a modified Sakai-Sugimoto model which
includes the open string tachyon stretching between the flavour branes
and antibranes. Taking the tachyon into account is essential for the
consistency of the setup. Our modification preserves the nice
geometric picture of chiral symmetry breaking of the Sakai-Sugimoto
model and at the same time relates chiral symmetry breaking to tachyon
condensation; the tachyon becomes infinitely large in the infrared
region where the joining of the flavour branes signals chiral symmetry
breaking.

We have shown that the tachyon condensate is essential to give the
goldstone bosons nonzero masses. We have identified parameters in the
tachyon field profile which correspond to the quark bare mass and
chiral condensate. We also briefly discussed different types of low
spin meson fluctuations. A more complete discussion with numerical
estimates for masses etc is under preparation.

There are several directions in which the present ideas can be
extended. It would be interesting to discuss this model at finite
temperature and describe the chiral symmetry restoration transition
and study the phase diagram in some detail. The connection with
tachyon condensation seems fascinating and a deeper understanding
would be useful. Finally, baryons have been discussed in the
Sakai-Sugimoto model. It turns out that they have a very small
size. This may change in the presence of the tachyon. This is because
in the presence of the tachyon, the flavour energy momentum tensor is
concentrated far away from the infrared region where the branes
meet. It would be very interesting to investigate whether this effect
actually makes a difference to the baryon size.

\gap5

%\noindent{\bf Acknowledgments} 

%\newpage

%\appendix

%\section{\label{solution}...........}

%\subsection{\label{large-U}Solution for large $U$}

\newcommand{\sbibitem}[1]{\bibitem{#1}}

\end{document}